\documentclass[prd,showpacs,superscriptaddress,twocolumn,
floatfix,preprintnumbers,altaffilletter,nofootinbib]{revtex4}
\usepackage[utf8]{inputenc}
\usepackage{longtable}
\usepackage{graphicx}
\usepackage{amsmath,amssymb}
\usepackage{color}
\usepackage{units}
\usepackage{epstopdf}
\usepackage{hyperref}
\usepackage{multirow}

\newcommand{\beq}{\begin{equation}}
\newcommand{\eeq}{\end{equation}}
\newcommand{\bdm}{\begin{displaymath}}
\newcommand{\edm}{\end{displaymath}}

\DeclareMathOperator{\dist}{dist}
\DeclareMathOperator{\leakyReLU}{leakyReLU}

\definecolor{Gray}{gray}{0.9}

\graphicspath{{./plots/}}
\begin{document}


\title{Classifying the unknown: discovering novel gravitational-wave detector glitches using similarity learning}

\author{S~Coughlin}
\email{scottcoughlin2014@u.northwestern.edu}
\affiliation{Physics and Astronomy, Cardiff University, Cardiff, CF10 2FH, UK}
\affiliation{Center for Interdisciplinary Exploration \& Research in 
(CIERA), Northwestern University, Evanston, IL 60208, USA}

\author{S~Bahaadini}
\affiliation{Electrical Engineering and Computer Science, Northwestern University, Evanston, IL, 60201, USA}

\author{N~Rohani}
\affiliation{Electrical Engineering and Computer Science, Northwestern University, Evanston, IL, 60201, USA}

\author{M~Zevin}
\affiliation{Center for Interdisciplinary Exploration \& Research in 
(CIERA), Northwestern University, Evanston, IL 60208, USA}

\author{O~Patane}
\affiliation{Department of Physics, California State University Fullerton, Fullerton, CA, 92831, USA}

\author{M~Harandi}
\affiliation{School of Information Studies, Syracuse University, Syracuse, NY, 13210, USA}

\author{C~Jackson}
\affiliation{School of Information Studies, Syracuse University, Syracuse, NY, 13210, USA}

\author{V~Noroozi}
\affiliation{Department of Computer Science, University of Illinois at Chicago, IL, 60607, USA}

\author{S~Allen}
\address{Adler Planetarium, Chicago, IL, 60605, USA}

\author{J~Areeda}
\affiliation{Department of Physics, California State University Fullerton, Fullerton, CA, 92831, USA}

\author{M~Coughlin}
\affiliation{Division of Physics, Math, and Astronomy, California Institute of Technology, Pasadena, CA 91125, USA}

\author{P~Ruiz}
\affiliation{Electrical Engineering and Computer Science, Northwestern University, Evanston, IL, 60201, USA}

\author{C~P~L~Berry}
\affiliation{Center for Interdisciplinary Exploration \& Research in 
(CIERA), Northwestern University, Evanston, IL 60208, USA}

\author{K~Crowston}
\affiliation{School of Information Studies, Syracuse University, Syracuse, NY, 13210, USA}

\author{A~K~Katsaggelos}
\affiliation{Electrical Engineering and Computer Science, Northwestern University, Evanston, IL, 60201, USA}

\author{A~Lundgren}
\affiliation{Institute of Cosmology and Gravitation, University of Portsmouth, Portsmouth, United Kingdom}

\author{C~{\O}sterlund}
\affiliation{School of Information Studies, Syracuse University, Syracuse, NY, 13210, USA}

\author{J~R~Smith}
\affiliation{Department of Physics, California State University Fullerton, Fullerton, CA, 92831, USA}

\author{L~Trouille}
\address{Adler Planetarium, Chicago, IL, 60605, USA}

\author{V~Kalogera}
\affiliation{Center for Interdisciplinary Exploration \& Research in 
(CIERA), Northwestern University, Evanston, IL 60208, USA}

\date{\today}

\begin{abstract}
The observation of gravitational waves from compact binary coalescences by LIGO and Virgo has begun a new era in astronomy.
A critical challenge in making detections is determining whether loud transient features in the data are caused by gravitational waves or by instrumental or environmental sources.
The citizen-science project \emph{Gravity Spy} has been demonstrated as an efficient infrastructure for classifying known types of noise transients (glitches) through a combination of data analysis performed by both citizen volunteers and machine learning.
We present the next iteration of this project, using similarity indices to empower citizen scientists to create large data sets of unknown transients, which can then be used to facilitate supervised machine-learning characterization.
This new evolution aims to alleviate a persistent challenge that plagues both citizen-science and instrumental detector work: the ability to build large samples of relatively rare events.
Using two families of transient noise that appeared unexpectedly during LIGO's second observing run (O2), we demonstrate the impact that the similarity indices could have had on finding these new glitch types in the Gravity Spy program.
\end{abstract}

\maketitle

\section{Introduction}\label{sec:intro}

The recent detection of gravitational waves (GWs) from the inspiral and merger of binary black holes \cite{AbEA2016a,AbEA2016e,AbEA2017a,AbEA2017c,Abbott:2017gyy,LIGOScientific:2018mvr} and binary neutron stars \cite{AbEA2017b,LIGOScientific:2018mvr} by the ground-based interferometric detectors Advanced LIGO~\cite{aligo} and Advanced Virgo~\cite{avirgo} has begun the era of GW astronomy. 
The analysis of GW data is complicated by the presence of noise transients of both instrumental and environmental origin known as \emph{glitches}. 
These noise transients are caused by a wide variety of phenomena, including up-conversion of ground motion into the optical system, laser power fluctuations, and magnetic fields at the local site \cite{Nuttall20170286}; 
however, many persistent noise transients have no known cause, and are not coincidentally witnessed by any auxiliary monitoring channel.
Glitches can impact the detection of signals, as they can be confused with astrophysical signals with short durations or significant theoretical uncertainties, and, if occurring the same time as a GW (as was the case for the first binary neutron star observation \cite{AbEA2017b}), they make it more challenging to accurately infer the properties of the astrophysical source \cite{Pankow:2018qpo,Powell:2018csz}. 
Over the next decade \cite{Aasi:2013wya}, sensitivity improvements to the GW detectors are expected due to increased laser power, improved optical coatings, and improved seismic isolation.
These improvements will also lead to an altered sensitivity to instrumental and environmental noise transients, as well as to changes in the character of some types of glitches \cite{ZeCo2017}.
As the Advanced LIGO and Advanced Virgo detectors evolve, glitches will continue to remain a challenge for GW analysis. 

Significant efforts are made to characterize and identify these noise transients so that the times that they are present can be removed from analyzed data, and, if possible, their causes eliminated altogether.
This can be accomplished through the use of algorithms that use auxiliary sensors placed in and around the GW detectors to identify those sensors that are highly correlated witnesses with a particular type of glitch \cite{SmAb2011, iDQ, Karoo}.
The idea is to identify activity in witness sensors that can be attributed to the glitch, thereby requiring the removal of as little GW data as possible while also successfully removing the classes of data transients.
GW signals will not appear in these auxiliary sensors, so vetoing transients based on auxiliary sensor data does not erroneously remove true GW signals. 
An essential first step in diagnosing the causes of glitches is identifying them as a feature in the data.

There has been significant recent interest in the application of machine learning (ML) to glitch identification \cite{BiBl2013,PoTr2015,PoTo2017,MuAb2017,GeSh2017,RaCu2018}, to the identification of correlated witness channels \cite{iDQ, Karoo}, and to GW searches \cite{LiYu2017,GeHu2017,GaWi2017}. 
Though the classification of glitches through ML approaches has shown promise, these approaches suffer from some  shortcomings. 
First, the supervised ML methods, where previously known classes of transients are given as a training set to the algorithm, have no immediate way to identify other classes also present in the data. 
Alternatively, unsupervised ML methods, where the algorithm seeks to learn the discriminative features of the data set in order to create its own classes or clusters of similar data, have the downside of decoupling the analysis from the understanding of how known classes relate to the detector. 
Moreover, unsupervised methods inevitably suffer from the need to validate the self-identified classes, as the clusters are far from exclusive because the features the algorithm learns from the unlabeled data set are not discriminative enough.
Both supervised and unsupervised ML techniques have merits, but neither is a perfect solution.

In an effort to address the glitch classification problem, we previously introduced \emph{Gravity Spy} \cite{ZeCo2017}.\footnote{\href{https://www.gravityspy.org}{www.gravityspy.org}}
This combines the crowd-sourcing power of citizen science with the rapid classification ability of ML  \cite{BaRo2017} to support the characterization of glitches in GW data.
Gravity Spy is hosted on Zooniverse, a leading online platform that has enabled over 1.5 million citizen scientists to analyze scientific data.
Gravity Spy users are asked to classify time--frequency plots depicting glitches into one of a number of classes.  
The large number of people supporting this work provide data sets of known glitch classes, which are then used as training sets for the ML algorithms.
The ML algorithms can then rapidly classify the entire data set of known glitches. 
These statistically pure data sets are then used for the purpose of long term trend studies as well as targeted auxiliary channel follow-up, e.g comparing humidity at the detector with the rate of the blip glitch \cite{bliphumidity}. 
Although the classification and verification of known classes has proven effective in Gravity Spy, it remains challenging to collect sufficient numbers of novel glitches to identify new classes. 
We note that unsupervised clustering has been shown to classify new types of glitches, e.g., Reverse Chirp, shown in Figure C2 of \cite{GeSh2017}.

To solve this problem we employ techniques from transfer learning.
Transfer learning applies the knowledge from a labeled data set to an unlabeled data set with different features.
In this case, we are interested in transferring the knowledge of what makes the known glitch classes in Gravity Spy similar and different from each other to the domain of images that do not belong to any known class.
Although this method proves useful in helping the algorithm extract more discriminating features that make for cleaner clustering of the unlabeled data, they are still too inclusive to confidently contain a single new class of glitch.
Therefore, combining the feature space obtained through transfer learning techniques with human controlled clustering of this space may prove the most effective way to rapidly identify new glitch classes.

We introduce a new method for the rapid identification of novel transients that combines techniques within the field of transfer learning with the crowd-sourcing power of Gravity Spy.
In Section~\ref{sec:similarity-learning}, we discuss the specifics of our transfer learning algorithm. 
We then discuss the new proposed infrastructure for Gravity Spy in Section~\ref{sec:gs}. 
In Section~\ref{sec:results}, we highlight the impact the proposed methodology could have had on discovering two sets of new LIGO glitches from the second observing (O2) run. In Section~\ref{sec:results-part2}, we summarize the impact of different settings of the transfer learning algorithm on the discriminative ability of the feature space.
In Section~\ref{sec:conclusions}, we discuss future iterations of the Gravity Spy project and its role in GW detector characterization.

\section{Transfer Learning}\label{sec:similarity-learning}

\begin{figure*}[t]
 \includegraphics[width=0.9\textwidth]{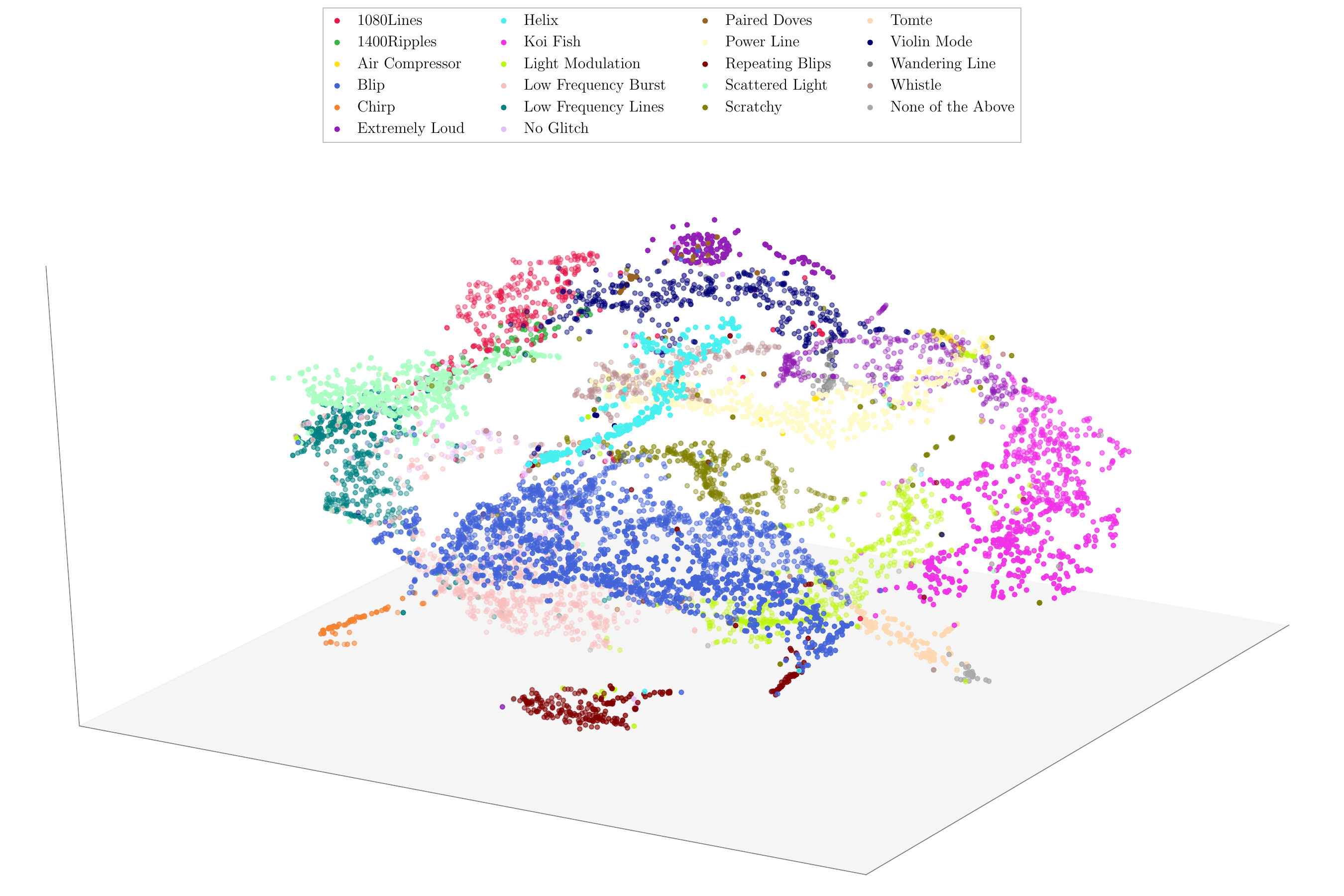}
 \caption{
Visual representation of the training set in the DIRECT feature space using the $t$-distributed Stochastic Neighbor Embedding ($t$-SNE) statistic. This metric is purely designed to project groups of samples in the $N$-dimensional feature space into $3$ dimensions and has no physical meaning.}

 \label{fig:trainingset}
\end{figure*}

Transfer learning applies knowledge obtained from a model that was trained on one data set to another data set.
Specifically, for our method, we hope to transfer knowledge about what makes the spectrograms of the known Gravity Spy classes similar and different from each other to the unlabeled Gravity Spy glitches.
We anticipate that this knowledge will enable a better clustering of these glitches that will lead to the discovery of new classes.

To accomplish this goal, we must first train an algorithm designed to model the similarity and differences between images on the known set of Gravity Spy glitches.
For this analysis, we use the transfer-learning algorithm DIRECT \cite{DBLP:journals/corr/abs-1805-02296} to quantify similarity between Gravity Spy images.
In short, this algorithm solves for a nonlinear embedding function $f_{\theta}$, i.e.\ the discriminative feature space, by using a deep neural network.
Using pairs of labeled images as input, the neural network is trained by solving for the $f_{\theta}$ that minimizes the function 
\begin{eqnarray}
\label{eq:obj}
\mathcal{L} & = & {} \sum_{i=1}^{N}l(y^i,x_1^i,x_2^i) \nonumber  \\ 
 & = & {} \sum_{i=1}^{N} y^i  {\dist\left(f_{\theta}(x_1^i), f_{\theta}(x_2^i)\right)}  \\*
 & & {} + (1-y^i) \max \left\{0, m-{\dist\left(f_{\theta}(x_1^i), f_{\theta}(x_2^i)\right)}\right\} \, . \nonumber
\end{eqnarray}
\noindent
Here $N$ is the number of training pairs; $x_1^i$ and $x_2^i$ are the first and second items of the $i$-th pair; $y^i$ is the binary label of the $i$-th pair, which is one when the two items of the pair belong to the same class and zero when they belong to different classes; $\dist$ is a distance function (such as Euclidean or cosine), and $m$ is the margin that is used to bound the distance between the items of pairs from different classes. 
A convolutional neural network models the nonlinear function $f_{\theta}$ by adding a fully connected dense layer onto the pre-trained VGG16 network \cite{DBLP:journals/corr/SimonyanZ14a}.
The VGG16 network consists of $13$ convolutional layers and $2$ fully connected layers and was pre-trained on the ImageNet \cite{imagenet_cvpr09} database of images.
We use the cosine distance metric
as our distance function,\footnote{The cosine distance between two vectors $\vec{a}$ and $\vec{b}$ is $1-\vec{a}\cdot\vec{b}/(|\vec{a}|\,|\vec{b}|)$.}
and to train the model we use the Gravity Spy training set described in \cite{BAHAADINI2018172}.
Each glitch is portrayed as four spectrograms with different temporal durations. These are generated using gwpy \cite{gwpy}.
We take these four images to create a single merged image for each glitch, identical to the input currently used for the convolutional neural network classifications in Gravity Spy \cite{BAHAADINI2018172}.
By propagating through the DIRECT network described above, the pixel data of the input image is mapped to a smaller, $200$ dimensional feature space. 
The dimension of the feature space is fixed at $200$ based on Fig.~2 of \cite{DBLP:journals/corr/abs-1805-02296}.
In Figure~\ref{fig:trainingset} (cf.\ Fig.~3 of \cite{BAHAADINI2018172}) we show a visual representation of the training set in the DIRECT feature space using the $t$-distributed stochastic neighbor embedding statistic ($t$-SNE) \cite{vandermaaten2008visualizing}. 
As can be seen  samples from the same glitch class are put closer to each other while samples of different classes are far from each other.
Such a property is called \emph{discriminative} feature representation.
Having trained this model on the known set of glitches, we can now apply it to the unlabeled glitches so that in this new, more discriminative feature space we can cluster similar images together and find new classes.

Having established a means of clustering glitches, in the next section, we describe how we use the discriminative feature space obtained using DIRECT on the Gravity Spy data set to empower volunteers to build large data sets of unknown glitches.

\section{Identifying Novel Glitches}\label{sec:gs}

\begin{figure*}[t]
 \includegraphics[width=0.9\textwidth]{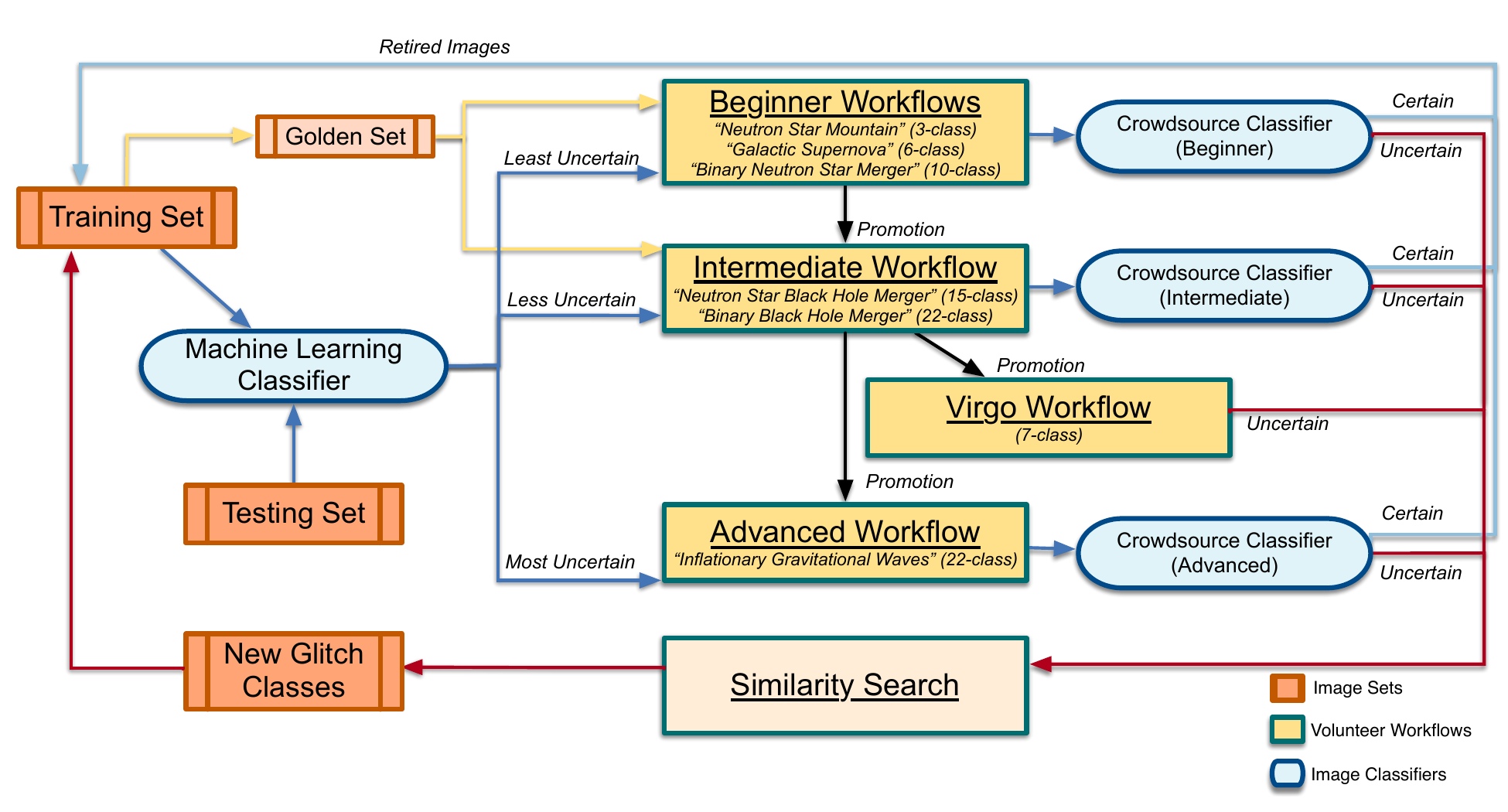}
 \caption{
New infrastructure proposal for Gravity Spy. This design differs from that described in \cite{ZeCo2017} by facilitating the direct follow-up of single examples of unknown transients through the similarity search algorithm. This is in contrast to the reliance on the None-of-the-Above classifications for filtering out novel glitches from the data set.
 }
 \label{fig:similarity-workflow}
\end{figure*}

In our previous work, we relied on the None-of-the-Above classification to identify glitches from previously unknown classes, and the volunteer \emph{Talk} forum (a thread of comments on the image from other volunteers) to consolidate examples in order to develop training sets and add new classes to the supervised model.
As the volume of data increases, this design will prove ineffective as a user would have to go through too many classifications before seeing multiple examples of a novel glitch.

To this end, we are introducing a similarity search tool to the Gravity Spy infrastructure.
This tool uses the feature space output for every Gravity Spy glitch, which is obtained using the DIRECT algorithm described in Section~\ref{sec:similarity-learning}, to enable users to take a single example of a glitch and perform a search querying for glitches in the data which are morphologically similar to the example.
In order to determine similarity, we subtract the same cosine distance metric described in Section~\ref{sec:similarity-learning} by one.
In this case, identical images have a distance metric of $1$, while orthogonal images, that is the two images share no common components in DIRECT feature space, images have a distance metric close to $0$.
This metric has the advantage of being efficient to evaluate and not being affected by the overall scaling of the input vectors.
Specifically, this tool is introduced in the form a supplementary web service gravityspytools.\footnote{\href{https://gravityspytools.ciera.northwestern.edu/}{gravityspytools.ciera.northwestern.edu}}
The new infrastructure design is highlighted in Figure~\ref{fig:similarity-workflow}.

We anticipate that this tool, which outputs results in the form of \emph{Collections} (galleries of user-selected Gravity Spy glitches), will reduce the size of the data set to be searched for examples of a new class, such that building substantial training sets for unknown glitch types is manageable by a single user.
In the next section, we demonstrate how this tool could have proved useful in the rapid identification of two glitches classes that appeared in O2 which were \emph{not} previously included in the Gravity Spy classification.

\section{Results}\label{sec:results}
\begin{figure*}[t]
 \includegraphics[width=0.9\textwidth]{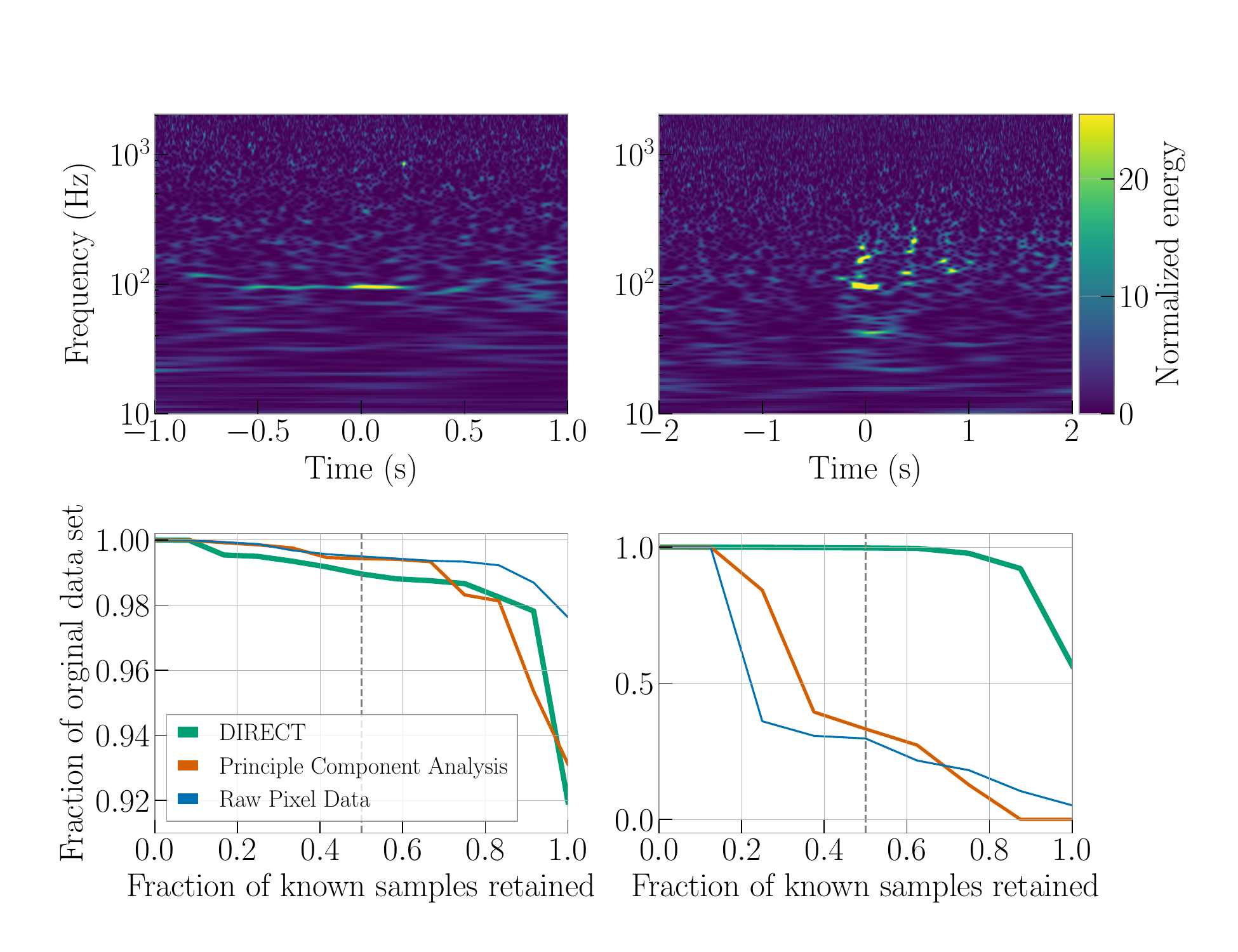}
 \caption{
\textit{Top}: Nominal examples of the Raven Peck (\textit{left}) and Water Jet (\textit{right}) glitches. 
\textit{Bottom left}: 
The fraction of known Raven Peck samples that have a higher similarity score than a given percentage of other data set samples when calculating similarity to a single known Raven Peck glitch.
For example, while retaining $50.0\%$ of known Raven Peck glitches, we can disregard about $99.0\%$  percent of the other data set samples, increasing the purity of the set to be examined by the user.
\textit{Bottom right}: 
Same for Water Jet glitches. Similarly, while retaining $50.0\%$ of known Water Jet glitches, we can disregard about $99.0\%$  percent of the other data set samples.
}
 \label{fig:raven-peck-water-jet-similarity-histogram}
\end{figure*}

We now highlight the application of the similarity search tool on data from O2.
Specifically, we assess the impact the similarity search tool could have had on the identification of two glitch classes that appeared during O2: the Water Jet \cite{water_jet} and the Raven Peck \cite{raven_peck} glitches.
The Water Jet glitch was caused by local seismic noise which resulted in loud bangs near the input optics, and the Raven Peck glitch was caused by ravens pecking on ice built up along vent lines transporting nitrogen outside of the detector \cite{Nuttall20170286}.
The resulting time--frequency morphologies of these glitches as they appear in the GW data channel can be seen in the top panel of Figure~\ref{fig:raven-peck-water-jet-similarity-histogram}.
These glitches occurred in the LIGO-Hanford detector.
The Raven Peck glitch is found in Gravity Spy data between 14 April 2017 and 9 August 2017, and the Water Jet glitch in data between 4 January 2017 and 28 May 2017.
Over these durations, there are a total of $13513$ and $26871$ Gravity Spy glitches, respectively.

We use these two glitches because they highlight the strengths and weaknesses of the DIRECT algorithm, and they allow us to emphasize the importance of incorporating crowd-sourcing methods into the identification of new glitches classes.
As DIRECT is a transfer learning algorithm, it is only able to employ concepts of similarity and difference learned from the training set to the unlabeled data.
If the distinguishing characteristic of a particular glitch is not also something present or extractable from the training set, it may be more difficult for the algorithm to find other examples easily.
This is demonstrated well here because the Raven Peck glitch is most prominently defined as a line feature which \emph{is} present in many of the Gravity Spy classes used in training (such as Power Line, Low Frequency Line and 1080 Line); 
on the other hand, the unique aspect of the Water Jet glitch is the subtle frequency decay that occurs after the initial pulse, which \emph{is not} obviously extractable from glitches in the training set.

We demonstrate the ability of the similarity search tool to organize testing images by their similarity to a queried glitch.
As was done in the DIRECT paper \cite{DBLP:journals/corr/abs-1805-02296}, we also compare finding similar images with DIRECT to more straightforward approaches such as using the raw pixel data or doing a Principle Components Analysis (PCA).
The bottom panel of Figure~\ref{fig:raven-peck-water-jet-similarity-histogram} shows the fraction of known samples that have a higher similarity score than a given percentage of the other data set samples.
For example, while retaining $50.0\%$ of known Raven Peck glitches, we can remove about $99.9\%$  percent of the other data set samples, increasing the purity of the set to be examined by the user.
For the same glitch, the raw pixel data approach and the PCA approach perform similarly with the raw pixels approach doing best at near $100.0\%$.
For Water Jet glitches, DIRECT also gives a similar performance retaining $50.0\%$ of known samples as it did for the Raven Peck.
However, if a retention rate of $100.0\%$ of the known samples is desired, the data set reduction rate for Raven Peck is $92.0\%$ compared to $55.0\%$ for the Water Jet glitch.
For this glitch, the methods of raw pixel data and PCA prove ineffective.
For a retention rate of $50.0\%$ only about $30.0\%$ percent of sample in the data set have lower similarity scores.
We believe these examples represent a challenging and less challenging task for the model, and in both cases DIRECT performs well, and the other approaches fail to be effective in the case of the Water Jet glitch.
We anticipate the reduction in the size of the original data set combined with the retention rate of similar samples to be significant enough that a single user can produce large data sets of novel glitches.

\subsection{Different Configurations}
\label{sec:results-part2}
To test the best training and setting configuration for DIRECT, we tried two different activation layers, $\tanh$ and $\leakyReLU$, for the custom fully connected layer that DIRECT adds to the VGG16 model, 
In addition, we varied the number of training rounds and the number of pairs of similar and dissimilar images that are drawn from the training set each time.
As training this model can be expensive because of the possible pairs of images one can produce from the training set, it is critical to understand what the minimal expense is that still produces an effective model.
To judge effectiveness, we quote the percentage reduction in the data set samples at which we still retain $50.0\%$ of other known Raven Peck and Water Jet glitches, respectively.
This value for each model is shown in Table~\ref{tab:results}.
The models using $\tanh$ as the activation layer perform worse than that with $\leakyReLU$.
We anticipate this is due to the restricted range allowed by the $\tanh$ activation layer, $[-1, 1]$, compared to that of $\leakyReLU$, $(-\infty, \infty)$.
Specifically, the distances away from each other similar and dissimilar images can be is restricted in the one case causing the discriminative feature space that is created to be less discerning than the other.
In terms of number of training pairs and rounds of training, it appears that increasing each does not lead to significantly improved results.
We anticipate this is due to the fact that most of the Gravity Spy classes are quite distinct from each other, and therefore using or drawing more pairs of images is unnecessary to produce a useful discriminate feature space representation of the data.
Therefore, this method can still be effective without an extremely costly training stage.

\begin{table}[]
\centering
 \begin{tabular}{c c c c c} 
 \hline
 Act. Layer & Train Rnds & 10,000 pr. & 50,000 pr. & 100,000 pr. \\ [0.5ex] 
 \hline\hline
 $\tanh$ & 30 & (0.93, 0.78) & (0.96, 0.92) & (0.94, 0.87) \\ 
  & 100 & (0.90, 0.85) & (0.93, 0.82) & (0.94, 0.72) \\
  & 200 & (0.93, 0.82) & (0.87, 0.80) & (0.93, 0.93) \\
  \hline
 $\leakyReLU$ & 30 & (0.96, 0.94) & (0.95, 0.92) & (0.95, 0.91) \\ 
  & 100 & (0.95, 0.91) & (0.96, 0.91) & (0.96, 0.94) \\
  & 200 & (0.95, 0.91) & (0.97, 0.92) & (0.96, 0.92) \\ 
 \hline
 \end{tabular}
 \label{tab:results}
 \caption{The fraction of the original data set with similar scores lower than the similarity score of $50.0\%$ of other known Raven Peck (left) and Water Jet glitches (right). Columns refer to different choices in the activation layer used in the dense layer of the model and the number of training rounds where each round draws a new set of X number of similar and dissimilar pairs. In bold is the configuration(s) that yielded the best reduction versus retention rate for both glitches.}
\end{table}

\section{Conclusions}\label{sec:conclusions}
We have described a novel extension of current GW data transient class identification combining the power of citizen scientists with the latest techniques in ML.
In the original paper, we allowed volunteers to classify glitches as None of the Above in order to identify individual subjects which belong to an unknown class.
Utilizing DIRECT, we have shown the ability to expedite the identification of new glitch classes compared with this original method, which will be important as we get new data from upcoming observing runs.

Using two noise transients from LIGO's O2 data, the Raven Peck and water Jet glitches, we demonstrated that DIRECT creates a discriminative feature space representation of the Gravity spy data set such that single examples of each glitch can efficiently lead to the discovery in the data set of other Raven Peck and Water Jet glitches. We compared DIRECT to simpler approaches such as using the raw pixel data or PCAs to find similar images and found that DIRECT produces either comparable or better results depending on the glitch.

There are a variety of plans for future related research.
For example, we can explore the use of other metrics to further the inter- and intra-class separation, thereby identifying separate classes that are otherwise improperly associated.
In addition, there are lessons learned that are applicable in other areas of astronomy, in line with  the on-going applications of unsupervised learning to large-scale astronomical surveys \cite[e.g.,][]{10.1093/mnras/sty3497, Khan:2018opv,2018MNRAS.476.3661D,2019PASP..131c8002M}.
For example, the Large Synoptic Survey Telescope (LSST) \cite{Ivezic2014}, an 8-meter class telescope being constructed on Cerro Pachon near La Serena, Chile which will take millions of images over its lifetime, identifying approximately 100,000 objects each night.
Although it is impossible for most of them to be analyzed by astronomers directly, citizen scientists can contribute to the monitoring, classifying, and annotating of spurious and surprising data. 
The expectation is that LSST, with its unprecedented field-of-view and rapid cadence, will discover a multitude of astrophysical phenomena, and can benefit from the ability to rapidly identify unique signals.
Fast transients which fade over a few days timescales, such as kilonova, which have not been identified in previous surveys are likely to be found and will constitute a new class of transient for this survey.

In general, the possibilities for further science education with citizen science initiatives, which place students on the edge of the scientific frontier, lie strongly in the identification of previously unknown phenomena.
Projects such as these create an environment where not all phenomena are known and understood, in contrast to textbook science lessons, and achieve a more realistic view of the wonder and challenges of science \cite{2018arXiv180500441D,Tsueng304766,PMID:27047583}.
For this reason, these initiatives help provide the foundation for further education in any scientific field, where the goal is to be able to follow a logical account of a problem to a solution, through the creation of a hypothesis, the taking of data, and the eventual explanation to understand the phenomena. 
Projects like this will have significant educational benefits and will impact the research projects both inside of LIGO and LSST and outside in numerous research groups conducting other astrophysical studies.
We anticipate that the combination of more novel ML techniques with web applications will continue to help with the efficacy of this work.

\acknowledgments
First, and foremost, we thank the many Gravity Spy participants that makes this work possible.
We thank Eliu Huerta, Alex Urban, and Patrick Sutton for their useful comments.
Gravity Spy is partly supported by the National Science
Foundation award INSPIRE 15-47880.
OP is supported by NSF award AST-1559694.
MC is supported by the David and Ellen Lee Postdoctoral Fellowship at the California Institute of Technology.
CPLB is supported by the CIERA Board of Visitors Research Professorship.
In addition, computing was provided by the LIGO Data Grid which is supported by the National Science Foundation Grants PHY-0757058 and PHY-0823459.
This work also used computing resources at CIERA funded by NSF PHY-1126812.
This paper has been assigned LIGO document
number LIGO-P1800352.

\bibliography{references}

\end{document}